\documentclass[footinbib,amsfonts,amsmath,amssymb,,floatfix,superscriptaddress,twocolumn]{revtex4-2}

\usepackage{graphicx}
\usepackage{hyperref} 
\usepackage[normalem]{ulem}
\usepackage{color}
\usepackage{dsfont}
\usepackage[utf8]{inputenc}
\DeclareUnicodeCharacter{2010}{-}
\usepackage{float}
\usepackage{wrapfig}
\usepackage{amsmath}
\usepackage{comment}
\usepackage{amssymb}
\usepackage{bbold}
\usepackage{times}
\usepackage{blindtext}
\usepackage[dvipsnames]{xcolor}
\DeclareMathOperator{\Tr}{Tr}
\usepackage{subfigure}

\begin{document}

\title{Controlling entanglement in a triple-well system of dipolar atoms}

\author{K. Wittmann W.}
\email{karinww@uol.com.br}
\affiliation{Instituto de F\'{i}sica da UFRGS, Porto Alegre, RS, Brazil}
\author{L. H. Ymai}
\affiliation{Universidade Federal do Pampa, Bagé, RS, Brazil.}
\author{B. H. C. Barros}
\affiliation{Instituto de F\'{i}sica da UFRGS, Porto Alegre, RS, Brazil}
\author{J. Links}
\affiliation{School of Mathematics and Physics - The University of Queensland, Brisbane, Australia.}
\author{A. Foerster}
\email{angela@if.ufrgs.br}
\affiliation{Instituto de F\'{i}sica da UFRGS, Porto Alegre, RS, Brazil}


\begin{abstract}
We study the dynamics of entanglement and atomic populations of ultracold dipolar bosons in an aligned three-well potential described by an extended Bose-Hubbard model. We focus on a sufficiently strong interacting regime where the couplings are tuned to obtain an integrable system, in which the time evolution exhibits a resonant behavior that can be exactly predicted.  Within this framework, we propose a protocol that includes an integrability breaking step by tilting the edge wells for a short time through an external field, allowing the production of quantum states with a controllable degree of entanglement. We analyze this protocol for different initial states and show the formation of highly entangled states as well as NOON-like states. These results offer valuable insights into how entanglement can be controlled in ultracold atom systems that may be useful for the proposals of new quantum devices.
\end{abstract}

\maketitle



\section{Introduction}
Quantum entanglement is a phenomenon discovered in the foundations of quantum physics that paved the way for a new era of technological advances. It represents non-local correlations between separate parts of a quantum system. 
As a resource, entanglement has been proven to be very useful for performing numerous tasks that face barriers in a classical setting,
finding broad applications in quantum information processing~\cite{shannon1948,nielsen2001quantum,Maldacena2016JHEP,Fadel2018,imany2019high,Niknam2020}, quantum teleportation~\cite{Gordon2006teleportation,HAFFNER2008,Bennett1993Teleporting, ma2012quantum,Zeilinger2015loophole}, 
and quantum metrology and sensing~\cite{pezze2018,Kaubruegger2021metrology,marciniak2022metrology}.

Entangled states are key ingredients in the proposals of protocols for the development of new quantum devices 
\cite{fogarty2013NOONinterf,Zinner2016quantum,fograty2019SciPost,jensen2019PRA,christensen2020,grun2022interfer,grun2022protocol}, and hence understanding the mechanisms for producing and controlling entangled states with a high degree of precision is of fundamental importance. In this context, the search for highly entangled states is the aim of many technological quantum applications \cite{Horodecki2009entanglement}, which can be exploited through different platforms.
Among these, ultracold atoms are especially interesting because they enable the manipulation of atoms arranged in optical potentials, with astonishing precision and versatility of the operating control~\cite{Bloch2005,Dumke_2016,mistakidis2022}.

In recent experiments on ultracold quantum gases, dipolar bosons are loaded into optical lattices to generate long-range dipole-dipole interaction (DDI), allowing access to fascinating novel quantum properties and phases~\cite{Trefzger_2011}.
The dynamics of such dipolar boson systems have been intensively studied and described, with good results, by an extended Bose-Hubbard model (EBHM)~\cite{Lahaye_2009,Petter2019}. 
One interesting feature of the EBHM with few bosonic modes is that the couplings of interactions can be tuned to achieve an integrable regime, which is particularly suited for the design of quantum devices. For instance, in \cite{tonel2020}, the conserved charge provided by integrability plays a crucial role when examining the quantum dynamics of a dipolar Bose-Einstein condensate (BEC) on a three-well aligned system, making it a potential candidate for constructing an atomic transistor \cite{Wittmann2018}. Other integrable quantum systems are being recently utilized to support the development of quantum technologies. These include quantum circuits created from transfer matrices \cite{Lucas2021} and those created through the star-triangle relation \cite{miao2023integrable}, central spin models for quantum sensors \cite{Villazon2022}, and 
the preparation of Bethe states on a quantum computer \cite{Dyke2021, VanDyke2021,Li2022, Sopena2022algebraicbethe}.

Here we consider an integrable triple-well model of dipolar bosons
and propose a protocol to create quantum states with controllable entanglement level. The control is realized by breaking the integrability for a short time and the resulting entanglement is characterized by the von Neumann entropy and correlation functions. We test the protocol for a range of different initial states, 
demonstrating how to produce highly entangled states 
as well as other important quantum states such as NOON-like states
\footnote{NOON state, belonging to the class of Schr{\"o}dinger cat states, is an ``all and nothing'' superposition of two different modes. See for example \cite{Dowling2002Rosetta,afek2010high,qi2022generating,grun2022protocol,qi2023NOONfloquet}.}

The paper is organized as follows. In section II, we describe the system and 
discuss the conditions for obtaining an effective description of the integrable system in the resonant regime. In section III, we analyse the dynamics of the system and the entanglement behaviour in the resonant regime. 
In Section IV, we propose a protocol for controlling entanglement by briefly tilting the edge sites of the system.
In sections V-VII, we analyse the action of the protocol on different initial states.  A discussion on interferometric applications of the protocol and details on the ground state structure are given in the appendices.  The conclusions are given in section VIII. 


\section{System description}

\begin{figure}[ht]
    \centering
        \includegraphics[width=1\linewidth]{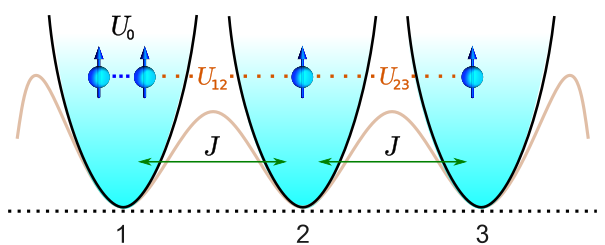}
    \caption{Schematic representation of the triple well system.
    The arrows represent the dipoles of atoms oriented along the direction of polarization. The coupling $J$ represents the hopping rate, $U_0$ characterizes on-site interactions, 
    while $U_{ij}$ characterizes the DDI between particles on different sites.}
    \label{fig:trap}
\end{figure}

We consider a system of dipolar atoms in an aligned triple-well potential described by
the following extended Bose-Hubbard model:
\begin{eqnarray}\label{h0}
H&=& \frac{U_0}{2}\sum_{i=1}^3 N_i(N_i-1) + \sum_{i=1}^3\sum_{j=1;j\neq i}^3\frac{U_{ij}}{2}N_i N_j\nonumber\\ 
&&-\frac{J}{\sqrt{2}}(a_1^\dagger a_2 + a_2^\dagger a_1 + a_2^\dagger a_3 + a_3^\dagger a_2), 
\end{eqnarray}
\noindent where $a_i$, $a_i^\dagger$ and $N_i=a_i^\dagger a_i$ are the bosonic annihilation, creation, and number operators of the well (or site) $i=1,2,3$, respectively. The coupling $J$ denotes the hopping rate of atoms between neighboring wells and $U_0$ and $U_{ij}=U_{ji}$ set the on-site and long-range interactions, respectively. The on-site interaction $U_0 = U_{sr} + U_{dd}$ results from short range interaction $U_{sr}$  and on-site dipole-dipole interaction (DDI) $U_{dd}$. The short-range interaction $U_{sr} \propto 4\pi\hbar^2 a/m$ is determined by the $s$-wave scattering length $a$ which is controlled through a magnetic field via Feshbach resonance and $m$ is the mass of atom. The on-site DDI $U_{dd}\propto \mu^2$ and long range interactions $U_{ij}\propto \mu^2$ obey an inverse cubic law whose strength is determined by the permanent magnetic dipole moment $\mu$ of dipolar atom considered and highly  depends on the geometry of potential trap and the polarization direction of dipoles~\cite{Lahaye_2009,baranov2008theoretical}. 
A schematic representation of this system is presented in Figure \ref{fig:trap}.

For the particular case when $U_{12}=U_{23}$ and $U_{13}=U_0$, the Hamiltonian given by Eq. \eqref{h0} is integrable \cite{Ymai2017} and can be reduced, up to a global constant, to
\begin{eqnarray}\label{hint}
H &=& U(N_1-N_2+N_3)^2\nonumber\\
&&-\frac{J}{\sqrt{2}}(a_1^\dagger a_2 + a_2^\dagger a_1 + a_2^\dagger a_3 + a_3^\dagger a_2),
\end{eqnarray}
where $N=N_1+N_2+N_3$ is the total number of particles and
$U=(U_0-U_{12})/4$ represents the effective interaction energy,
which can be experimentally tuned by controlling the polarization direction and depth of the well potential trap. 
A discussion on the feasibility of a physical realization of this system by means of Bose-Einstein condensates of dipolar atoms can be found in \cite{tonel2020}.
In this integrable case, the model can be formulated and solved using the quantum inverse scattering method and Bethe ansatz methods \cite{Ymai2017}. It acquires the additional conserved operator
\begin{eqnarray}
Q = \frac{1}{2}(N_1+N_3-a_1^\dagger a_3-a_3^\dagger a_1),
\end{eqnarray}
besides the Hamiltonian $H$ and the total number of particles $N$,
resulting in three independent conserved operators in an equal number of system modes.
The conserved charge $Q$ plays an important role in the {\it resonant regime}, characterized by the tunneling of atoms between the wells at the edges (labeled by $i = 1$ and $3$), 
while the number of particles in the middle well ($i=2$) remains approximately constant. 
This resonant behavior is a consequence of a second-order process that occurs in a relatively strong interaction regime \cite{Lahaye_2009,Lahaye2010}.
More specifically, when $|U(N-2l)/J| \gg 1$ and for the initial state
\begin{equation}
\vert \Psi_0 \rangle = |N-k-l,l,k\rangle, \label{is}
\end{equation}
where $l$ ($l=0,..., N$) and $k$ ($k=0,..., N-l$) represent the number of atoms initially at wells $2$ and $3$, respectively, 
the quantum dynamics of the Hamiltonian (\ref{hint}) can be well described by the effective Hamiltonian \cite{tonel2020},
\begin{eqnarray}
H_{\text{eff}} = \omega_l Q, \label{heff}
\end{eqnarray}
where constant $\omega_l$ is given by
\begin{equation}\label{freq}
\omega_l = \lambda_l J^2     
\end{equation}
with $\lambda_l= \frac{1}{4|U|}\left(\frac{l+1}{N-2l-1}-\frac{l}{N-2l+1}\right)$ depending on the initial number $l$ of bosons in the middle well.
The constant $\omega_l$ will play the role of the resonant tunneling frequency, with period $T_l=2\pi/\omega_l$. For the case where $l = 0$, let us simply denote it by $T\equiv T_0$. 

In the following sections, we first discuss the dynamical quantities that characterize the behavior of the system and provide information about its quantum entanglement. After that, we provide a protocol that briefly tilts wells 1 and 3 to control the entanglement of the quantum state.
Then, we analyze the effects of the protocol on different initial states.


\section{Dynamics of populations and entanglement}

We start by considering the dynamics of the system described above in the integrable and resonant regime, and for convenience, we set $\hbar = 1$. We focus on the time evolution of the average number of particles per well
\begin{equation}
   \langle N_i\rangle =\langle \Psi(t)\vert N_i\vert \Psi(t)\rangle
\end{equation}
and the von-Neumann entanglement entropy
\begin{equation}\label{eq:si}
    S_i(t)=-\Tr \left[\rho_i(t)\log\rho_i(t)\right]
\end{equation}
where the density matrix is defined as $\rho(t)=\vert \Psi(t) \rangle \langle \Psi(t) \vert$ and $\rho_i(t)$ is the reduced density matrix of site $i$ where the remaining subsystem is traced out. The von-Neumann entropy quantifies the bipartite entanglement between the site $i$ and the subsystem of other two sites. In the integrable regime, an initial state described by $\vert \Psi_0\rangle$ will evolve in time according to
\begin{equation}
\vert \Psi(t)\rangle = \mathcal{U}(t)\vert \Psi_0\rangle
\end{equation}
where $\mathcal{U}(t) \equiv e^{-i H t}$ is the time-evolution operator.
In what follows, we will use $|\Psi\rangle$ to refer to states obtained using Hamiltonian \eqref{hint}, 
and we will use the notation $|\widetilde{\Psi}\rangle$ with a tilde to denote analytic states obtained using the effective Hamiltonian \eqref{heff}, from which analytic results can be derived.
A comparison between the quantum states $|\Psi\rangle$ and $|\widetilde{\Psi}\rangle$ obtained for the same set of parameters will be quantified through the fidelity defined as $F= |\langle \Psi|\widetilde{\Psi}\rangle|^2$.
We will consider that the state $|\Psi\rangle$ theoretically approaches the analytic state $|\widetilde{\Psi}\rangle$ when $F > 0.95$.

For the case of initial state (\ref{is}), the state $|\widetilde{\Psi}(t)\rangle$ predicts that $\langle N_2\rangle = l$ remains constant,  while the atoms oscillate harmonically between sites $i=1$ and $i=3$, according to the expectation values given by
 \cite{tonel2020} 
\begin{equation}
\langle N_i\rangle =\frac{1}{2}\left[N-l+(2-i)(N-l-2k)\cos(\omega_l t)\right]. \label{N_i-eqs}
\end{equation}

\begin{figure}[ht]
    \centering
        \includegraphics[width=1\linewidth]{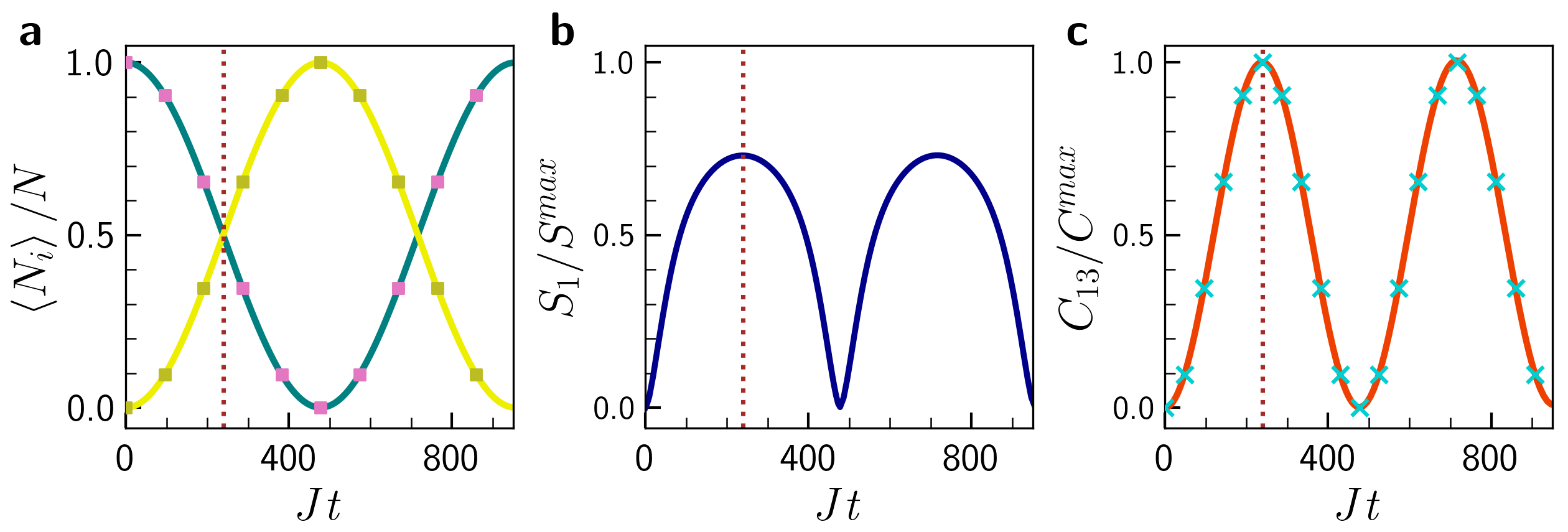}
    \caption{(\textbf{a}) Expectation value of $N_1/N$ (green line) and $N_3/N $ (yellow line) for the initial state $\vert \Psi_0\rangle = \vert 20,0,0\rangle$, $U=-2$ and $J=1$.  The marks represent the analytic expression \eqref{N_i-eqs}. (\textbf{b}) Time evolution of the entanglement entropy $S_1$ in units of $S_1^{max}=\ln(N+1)$. (\textbf{c}) Time evolution of correlation functions $C_{13}$ in units of $C_{13}^{max}=N/4$. The numerical simulation \eqref{eq:Cij} is represented by the solid line, while the marks results from the closed form expression. The vertical dotted lines mark the instant $t=T/4$, where $\langle N_1 \rangle = \langle N_3 \rangle$ and the values of $S_1$ and $C_{13}$ are maximum.}
    \label{fig:int}
\end{figure}

The expression above shows a maximum amplitude with oscillations of period $T=2\pi/\omega_0$ when all atoms are initially located in one of the edge wells (i.e., when $|\Psi_0\rangle = |N,0,0\rangle$ or $|\Psi_0\rangle = |0,0,N\rangle$ ) and an equilibrium with $\langle N_1\rangle = \langle N_3\rangle = N/2$ remaining constant when the edge wells initially have the same number of atoms (i.e., when $|\Psi_0\rangle=|l,0,l\rangle$, with $N=2l$). These two extreme cases will be the subject of our study later on.
In Figure \ref{fig:int} are shown some numerical results for the case $|\Psi_0\rangle = |N,0,0\rangle$ using the Hamiltonian \eqref{hint}.
Figure \ref{fig:int}-a shows the perfect agreement between the results of the numerical simulation and the expectation values given in (\ref{N_i-eqs}). 
In Fig. \ref{fig:int}-b the time evolution of the entanglement entropy $S_1$ presents a period of $T /2$ and its first maximum occurring at $t = T /4$, exactly when the populations reach an equilibrium with $\langle N_1\rangle=\langle N_3\rangle$. Nevertheless, despite the von-Neumann entanglement entropy is the most frequently used measure to quantify entanglement, it does not depend on any particular observable, making it difficult to perform a direct experimental measurement of its magnitude.
In order to generate signatures to indicate the formation of highly entangled states, in addition to enabling experimental measurements, we also evaluate the two-site correlation function defined as
\begin{equation}\label{eq:Cij}
  C_{ij}\equiv  |\langle N_i\rangle \langle N_j\rangle-\langle N_iN_j\rangle|.  
\end{equation}
Using the state $|\widetilde{\Psi}(t)\rangle$, we can derive the correlation function in the closed form $C_{13}= (N/4)\sin^2(t\,\omega_0)$, 
from which a maximum value is directly obtained at $t=T/4$. The agreement between numerical simulation of $C_{13}$ and its analytic formula can be seen in Fig. \ref{fig:int}-c. 
From Figures \ref{fig:int}-b and c it is clear that the maximum values of the two-site correlation functions and the entanglement entropy occur simultaneously. This result shows that the two-site correlation function is also able to reveal information about the quantum entanglement of the subsystem of wells.

It is worth noting that the entanglement entropy $S_2$ vanishes, since the state of site 2 remains constant in resonant regime, while $S_1=S_3\neq 0$ showing that bipartite entanglement is present only in the subsystem of sites 1 and 3.


\section{Protocol for quantum entanglement control}

\begin{figure}[h]
    \centering
                \includegraphics[width=1\linewidth]{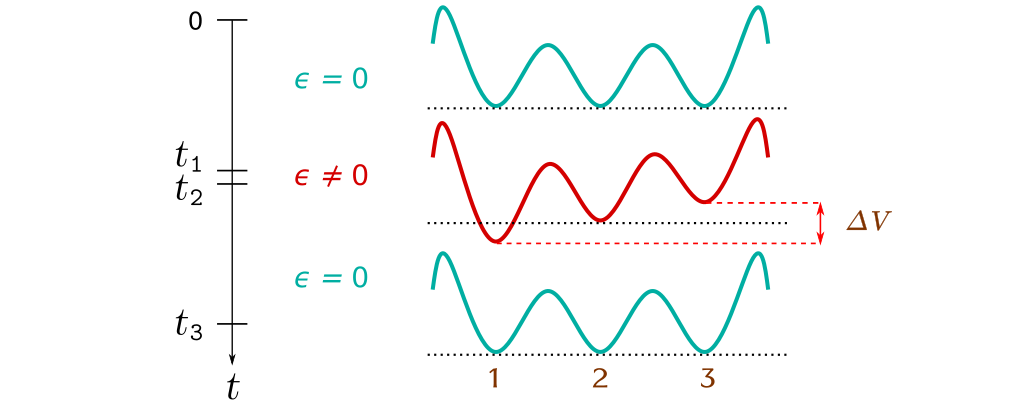}
        \caption{Schematic representation of the control protocol sequence. The tilt $\Delta V$ between the potentials of sites 1 and 3 is induced through an external field. The duration of the field allows the entanglement in the subsystem consisting of sites 1 and 3 to be controlled.}
    \label{fig:protocol}
\end{figure}

We now focus on establishing a protocol for generating and controlling maximally entangled states. The control of quantum entanglement can be achieved by tilting wells 1 and 3 through the action of an additional coherent light beam superimposed on the triple well system designed on an optical trap. During the presence of tilt on wells 1 and 3 the dynamics is governed by the Hamiltonian \cite{Wittmann2018} 
\begin{equation}\label{eq:Hepsilon}
    \mathcal{H}(\epsilon) = H + \epsilon(N_3-N_1),
\end{equation}
where $H$ is the integrable Hamiltonian \eqref{hint} and the parameter $\epsilon$ characterizes the energy offset between edge potential wells (see Figure \ref{fig:protocol}). 
Other properties of this Hamiltonian \eqref{eq:Hepsilon} can be found in \cite{tonel2020, Wittmann2022}.
Here we will examine the case where the tilt is introduced into the protocol as a short-duration square pulse just after the initial state evolves to the state with maximum correlation at $t=T/4$, as identified in the previous section. It will be seen that the amount of quantum entanglement is completely determined by the duration of the square pulse.

The full description of the protocol can be represented as follows
\begin{eqnarray}
|\Psi(t)\rangle = |\Psi_k(t)\rangle, \quad t_{k-1}\leq t\leq t_{k},\nonumber 
\end{eqnarray}
where the states for steps $k=1,2,$ and $3$ of the protocol are given sequentially by
\begin{eqnarray}
|\Psi_1(t)\rangle &=& \mathcal{U}(t-t_0,0)|\Psi_0\rangle, \nonumber\\
|\Psi_2(t)\rangle &=& \mathcal{U}(t-t_1,\epsilon)|\Psi_1(t_1)\rangle, \nonumber\\
|\Psi_3(t)\rangle &=& \mathcal{U}(t-t_2,0)|\Psi_2(t_2)\rangle. \nonumber
\end{eqnarray}
Here, $\mathcal{U}(t,\epsilon) \equiv e^{-i \mathcal{H}(\epsilon) t}$ is the time evolution operator. This sequence is depicted in Figure \ref{fig:protocol} below, illustrating the dependence on the parameter $\epsilon$.

In what follows, we continue to adopt the notation $|\Psi\rangle$  (without tilde) for states obtained using the Hamiltonian \eqref{eq:Hepsilon}, and $|\widetilde{\Psi}\rangle$  for analytic states obtained using the effective Hamiltonian $\mathcal{H}_{\text{eff}}(\epsilon) = H_{\text{eff}} + \epsilon(N_3-N_1)$ where $H_{\text{eff}}$ is given by \eqref{heff}. 


At the end of the whole process, the protocol generates the state
\begin{eqnarray}\label{eq:out}
|\Psi_{\text{out}}\rangle&\equiv &|\Psi_3(t_3)\rangle\nonumber\\
&=& \mathcal{U}(\Delta t_3,0)\mathcal{U}(\Delta t_2,\epsilon)\mathcal{U}(\Delta t_1,0)|\Psi_0\rangle,\\
\nonumber
\end{eqnarray}
where $\Delta t_k=t_k-t_{k-1}$, is the duration of $k$-th ($k=1,2,3$) step of the protocol. As mentioned earlier, we are assuming that $t_0 =0$, $\Delta t_1 = T/4$ and $\Delta t_2\ll \Delta t_{1,3}$ such that the breaking of integrability is the dominant effect in the second step of protocol. In the following sections, the action of the protocol on different initial input states will be investigated in detail.


\section{Input Fock state}
\begin{figure}
\centering
\includegraphics[width=17.8cm]{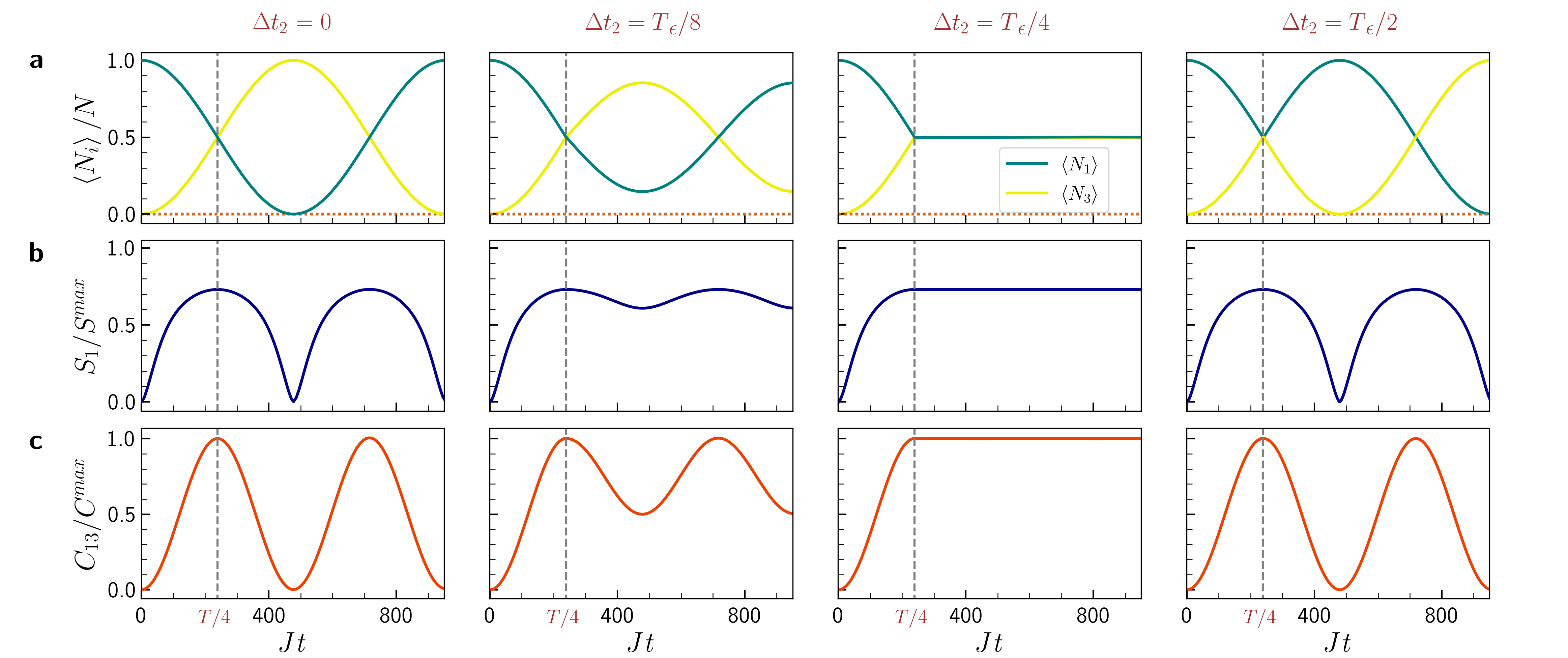}
    \caption{Time evolution of the expectation values (row \textbf{a}) $N_1/N$ (green line), $N_2/N$ (dotted line), $N_3/N$ (yellow line); entanglement entropy in units of $ S_1^{max} = \log(N+1)$  (row \textbf{b}) and, two-site correlation function in units of $C_{13}^ {max}=N/4$ (row \textbf{c}). Each column represents a different value of $\,\Delta t_2$. The first column represents the integrable case, where $\,\Delta t_2 = 0$. The other columns show the cases where $\, \Delta t_2 = T_\epsilon/8$,  $\Delta t_2 = T_\epsilon/4$ and $\, \Delta t_2 = T_\epsilon/2$, in sequence. In all cases, initial state $\vert \Psi_0\rangle= \vert 20,0,0\rangle$, $U=2$, $J=1$ and $\epsilon = 1$. The vertical dashed lines represent the instant $t=t_2$.}
\label{fig4}
\end{figure}

We start by first considering the case of a completely localized initial state given by $|\Psi_0\rangle=|N,0,0\rangle$. In Figure \ref{fig4}, it is shown the effect of the protocol on the dynamics for different values of duration of a square pulse, $\Delta t_2$, counted in units of period $T_\epsilon = 2\pi/\Omega_\epsilon$, where $\Omega_\epsilon=2\epsilon$.

In the first line of the Fig. \ref{fig4}, after the action of the square pulse ($t\geq t_2$),  it is shown the expectation value of the fractional population of sites $i=1,3$, which is given by

\begin{equation}
    \langle N_i\rangle/N = 1/2-(1-i/2)\sin[\omega_0 (t-t_2)]\cos \phi,
\end{equation}
and $\phi$ is a dimensionless parameter defined as
\begin{eqnarray}
\phi = 2 \epsilon \Delta t_2.\nonumber
\end{eqnarray}


We observe that the amplitude of the expectation values of $ N_i/N$ decrease gradually with increasing the pulse duration $\Delta t_2$ until the dynamics becomes stationary balanced for a long time at $\Delta t_2=T_\epsilon/4$ (or $\phi = \pi/2$) and completely reversed at $\Delta t_2 = T_\epsilon/2$ (or $\phi=\pi$). In the second line of Fig. \ref{fig4}, the range of values of entanglement entropy gradually decreases with increasing duration of the pulse, becoming stationary at its maximum value at $\Delta t_2 = T_\epsilon/4$. The dynamics of entanglement of the state $|\Psi(t)\rangle$ along the control process is also signaled in the third line of Fig. \ref{fig4} through the correlation function of sites 1 and 3.


\onecolumngrid

\newpage

\twocolumngrid

\begin{figure}[ht]
    \centering
                \includegraphics[width=1\linewidth]{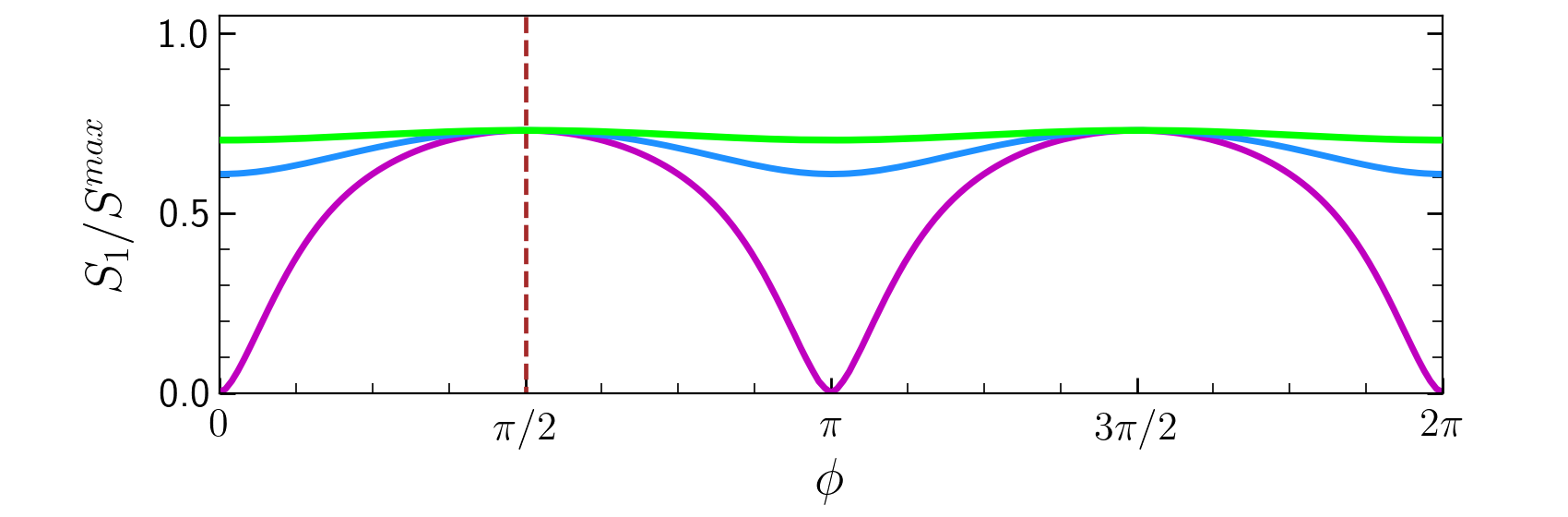}
    \caption{Entanglement entropy of state $|\Psi_{\text{out}}\rangle$ as function of $\phi\in [0, 2\pi]$ for $\Delta t_3=T/16$ (green) , $\Delta t_3=T/8$ (blue) and $\Delta t_3 = T/4$ (magenta), using the initial state $\vert \Psi_0\rangle = \vert 20,0,0\rangle$, $U=2$, $J=1$ and $\epsilon=1$. The dashed vertical line represents $\Delta t_2 = T_\epsilon/2$.}
    \label{fig:my_label3}
\end{figure}

In Figure \ref{fig:my_label3} we present the entanglement entropy of state $|\Psi_{\text{out}}\rangle$ as a function of $\phi$ for three time intervals $\Delta t_3 = T/16,\, T/8,\,T/4$. 
We observe that entanglement entropy can be controlled over a larger range of values at $\Delta t_3 = T/4$. Therefore, for fixed duration $\Delta t_3 = T/4$, the protocol predicts the following quantum state 
{\small
\begin{eqnarray}
|\widetilde{\Psi}_{\text{out}}(\phi)\rangle = \frac{\left[\sin(\phi/2)\, a_1^\dagger +\cos(\phi/2)\, a_3^\dagger\right]^N}{\sqrt{N!}}|0,0,0\rangle,
\label{output1}
\end{eqnarray}}

\noindent
where $|0,0,0\rangle$ is the vacuum state. From the above expression, the correlation function of sites 1 and 3 can be determined analytically as a function of parameter $\phi$ and it is given by $C_{13} =(N/4)\sin^2\phi$.
When performing the control within the interval $\Delta t_2\in[0,T_\epsilon]$,  the expression above shows that the maximized correlation $C_{13}^{max} = N/4$  occurs at $\phi = \pi/2\, (3\pi/2)$, when the state $|\Psi_{\text{out}}\rangle$ has the maximum entanglement entropy with all atoms into the (anti)symmetric coherent state with fidelity $F=0.99827\, (0.999509) $:
{\small
\begin{eqnarray}
|\widetilde{\Psi}_{\text{out}}\rangle = \frac{\left( a_1^\dagger \pm a_3^\dagger\right)^N}{\sqrt{2^N N!}}|0,0,0\rangle.\nonumber
\end{eqnarray}}
If all atoms are initially loaded into site 3 (i.e., $|\Psi_0\rangle = |0,0,N\rangle$), the states with maximum correlation are generated with symmetry reversed compared to the case where $|\Psi_0\rangle = |N,0,0\rangle$. In the next section, we consider the case where initially both sites 1 and 3 have the same number of atoms.


\section{Twin-Fock input state}

\begin{figure}[ht]
\centering
\includegraphics[width=1\linewidth]{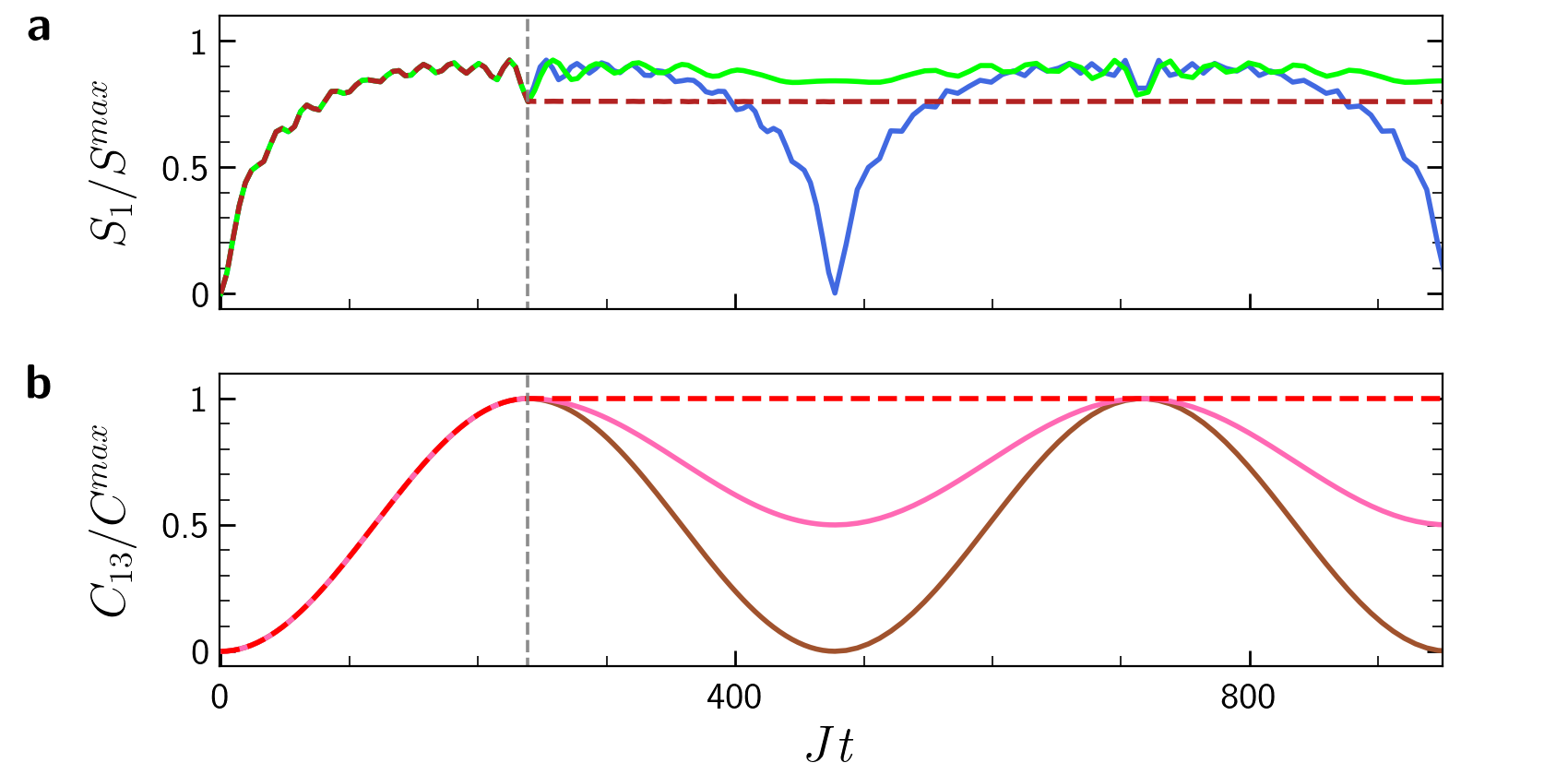}
\caption{Entanglement entropy (\textbf{a}) for  $\,\Delta t_2 = 0$ (blue line), $\,\Delta t_2 =T_\epsilon/8$ (green line) and $\,\Delta t_2 = T_\epsilon/4$ (dashed line); and correlation (\textbf{b}) for  $\,\Delta t_2 = 0$ (brown line), $\,\Delta t_2 =T_\epsilon/8$ (pink line) and $\,\Delta t_2 = T_\epsilon/4$ (dashed line). For all the cases, it was used  the initial state $\vert \Psi_0\rangle = \vert 10,0,10\rangle$, $U=2$, $J=1$, and $\epsilon = 1$. The vertical dashed lines represent the instant $t=t_2$.} 
\label{fig:my_label4}
\end{figure}

In this section we investigate the quantum entanglement control for the case of the initial twin-Fock state in the sites 1 and 3, given by $|\Psi_0\rangle = |l,0,l\rangle$, for which $N=2l$ and the expectation values $\langle N_1\rangle=\langle N_3\rangle=l$ remain constant under integrable time evolution in the resonant regime. 
Figure \ref{fig:my_label4} presents the dynamics of the entanglement entropy of state $|\Psi(t)\rangle$ for three different duration of a square pulse. 
Again, for $\Delta t_2 = T_\epsilon/4$, the entanglement entropy is stationary.

\begin{figure}[ht]
    \centering
                    \includegraphics[width=1\linewidth]{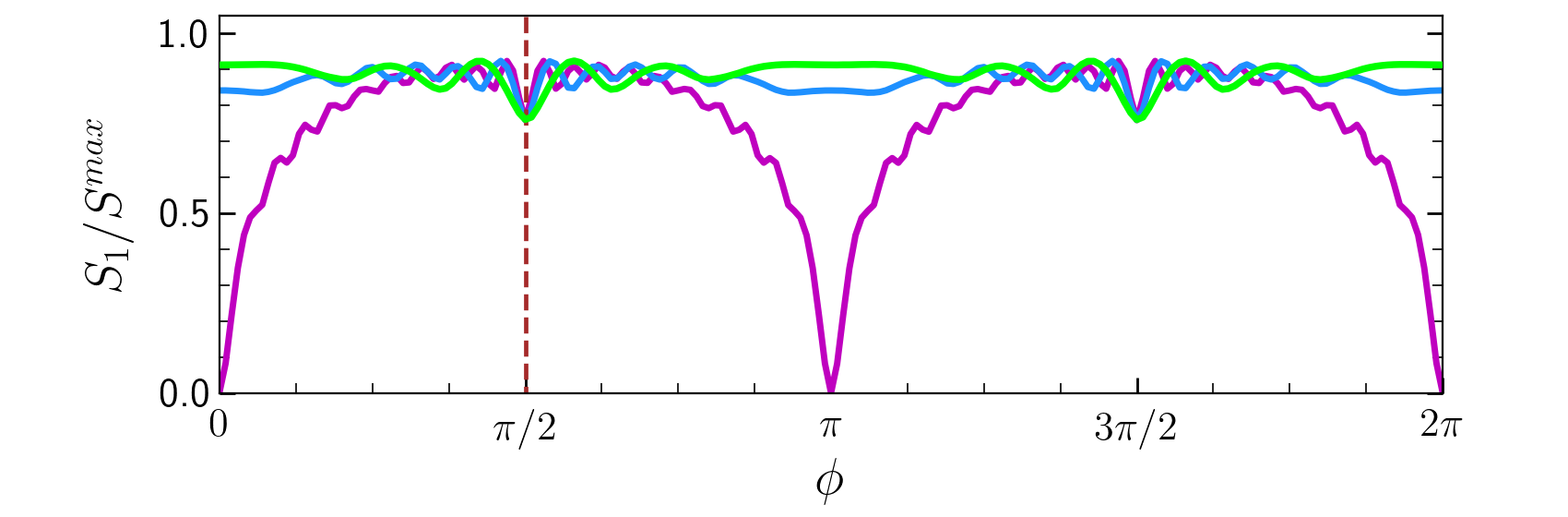}
    \caption{Entanglement entropy of state $|\Psi_{\text{out}}\rangle$ as function of $\phi\in [0, 2\pi]$ for $\Delta t_3 = T/4$ (magenta), $\Delta t_3=T/8$ (blue) and $\Delta t_3=T/16$ (green), using the initial state $\vert \Psi_0\rangle = \vert 10,0,10\rangle$, $U=2$, $J=1$, $\epsilon=1$ and $\Delta t_1=T/4$. The dashed vertical line represents $\Delta t_2 = T_\epsilon/2$.} 
    \label{fig:my_label5}
\end{figure}

The Figure \ref{fig:my_label5} shows the entanglement entropy $S_1$ as function of dimensionless parameter $\phi$ and three time intervals $\Delta t_3 = T/16,\, T/8$, and $T/4$.  
In this case, the output state $|\Psi_{\text{out}}\rangle$ presents high entanglement entropy with a small dip at $\Delta t_2 = T_\epsilon/4$ in its signature and for $\Delta t_3=T/4$ the state predicted by our protocol is given by 
{\small
\begin{eqnarray}\label{output2}
|\widetilde{\Psi}_{\text{out}}(\phi)\rangle = \frac{\left(2\cos\phi\,a_1^\dagger a_3^\dagger +\sin\phi[(a_1^\dagger)^2-(a_3^\dagger)^2]\right)^{l}}{2^{l}l!}|0,0,0\rangle.\nonumber\\
\end{eqnarray}}

The above state allows determining analytically the correlation function of sites 1 and 3 as a function of parameter $\phi$, given by
\begin{eqnarray}
C_{13} =\frac{l(l+1)}{2}\sin^2\phi.\nonumber
\end{eqnarray}
In particular, for $\phi=\pi/2$ and $\phi=3\pi/2$, the correlation achieves its maximum value $C_{13} = l(l+1)/2$ and the state $|\widetilde{\Psi}_{\text{out}}\rangle$ is highly entangled with the respective fidelities $F=0.998315$ and $F=0.998771$, given by (up to global phase)
\begin{eqnarray}
|\widetilde{\Psi}_{\text{out}}\rangle = \frac{\left[(a_1^\dagger)^2- (a_3^\dagger)^2\right]^{l}}{2^{l}l!}|0,0,0\rangle.\nonumber
\end{eqnarray}
The above state shows that the protocol acts on the initial twin-Fock state by performing a discrete Fourier transform on the modes 1 and 3 defined as $a_{1(3)}^\dagger \to (a_1^\dagger\pm a_3^\dagger)/\sqrt{2}$~\cite{islam2015measuring},
which leads to a quantum state with only an even number of particles at sites 1 and 3. This result can be interpreted as a destructive interference process on the odd number of particles, similar to the well-known Hong-Ou-Mandel (HOM) effect~\cite{rarity1990HOMfoton,lewis2014metterHOMproposal}. 

It is worth noting that, in the resonant regime, the time evolution operators $\mathcal{U}(T/4,0)$ and $\mathcal{U}(\Delta t_2,\epsilon)$ used to generate the output state $|\Psi_{\text{out}}\rangle$ play an analogous role of the $50:50$ beam-splitter and phase shifter operations in a Mach-Zehnder (MZ) interferometer \cite{Yurke1986interferometers, berrada2013integrated}
This shows the protocol is capable of performing interferometric operations in which the phase estimation sensitivity depends on the choice of the initial state and the observable to be detected. (see Appendix A). 


\section{Entangled input state}\label{input_cs}
In the previous sections, we considered a class of non-entangled initial states in which the state of well 2 remains constant over time, and therefore remains disentangled from the rest of the system. 
Now we will consider an entangled initial state in which quantum entanglement between well 2 and the subsystem composed of the other two wells is also manifest. To this end, let us analyze the effect of the protocol on the initial state defined as
\small{
\begin{eqnarray}\label{NLS}
|\Psi_0\rangle = \frac{1}{\sqrt{2}}|0,N,0\rangle+\frac{1}{\sqrt{2N!}}\left(\frac{a_{1}^\dagger+a_3^\dagger}{\sqrt{2}}\right)^N|0,0,0\rangle. 
\end{eqnarray}}

The above state has a NOON-like state (NLS) structure, in the sense it is a superposition between the state with all particles in well 2 and the state with all atoms in the subsystem of wells 1 and 3. However, the state of the subsystem of well 1 and 3 has all particles into a coherent state $|\text{CS}\rangle = \frac{1}{\sqrt{N!}}\left(\frac{a_{1}^\dagger+a_3^\dagger}{\sqrt{2}}\right)^N|0,0,0\rangle$. The motivation for its study is directly related to the ground state of the integrable Hamiltonian (see Appendix B).

\begin{figure}[ht]
    \centering
                \includegraphics[width=8.5cm]{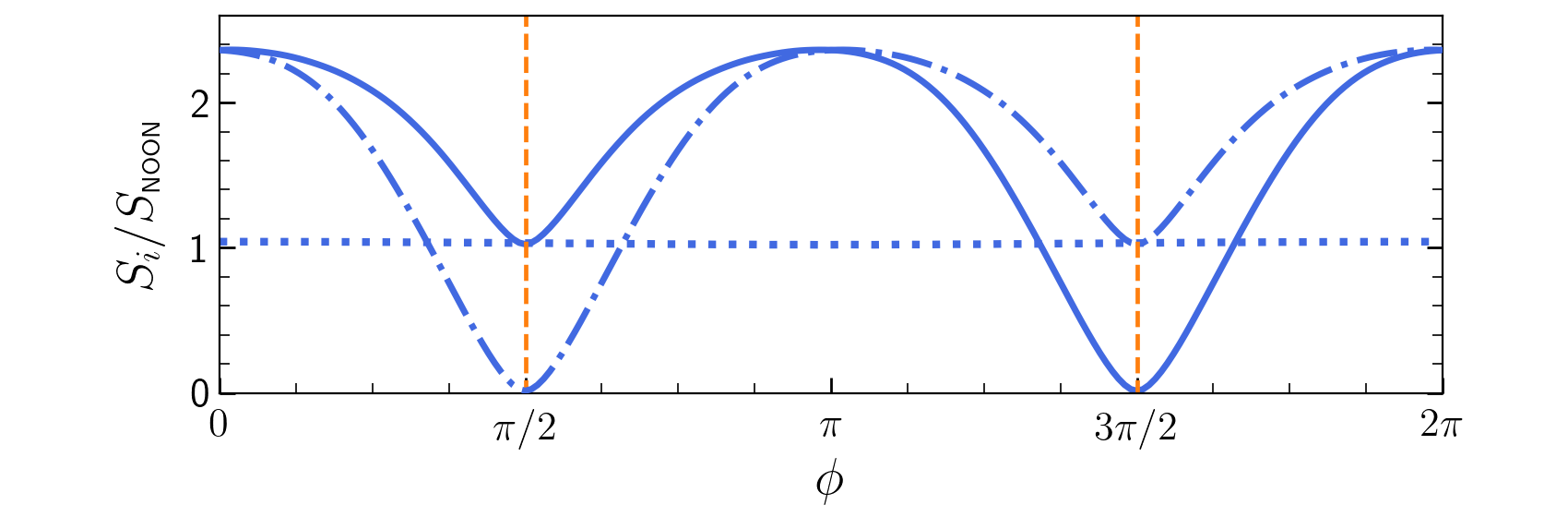}
    \caption{Entanglement entropies (in units os $S_{\text{NOON}}=\log 2$) $S_1$ (solid line), $S_2$ (dot line) and $S_3$ (dot-dashed line) of state $|\widetilde{\Psi}_{\text{out}}\rangle$ as function of $\phi\in [0, 2\pi]$, for $N=10$ and $U=-1.3$.} \label{fig:my_label7}
\end{figure}

Now, considering the case of initial state \eqref{NLS}, the effective Hamiltonian is still given by (\ref{heff}) with $l = 0$,  since  $\omega_N=\omega_0$. Then, the protocol predicts the following quantum state 
\begin{eqnarray}
|\widetilde{\Psi}_{\text{out}}(\phi)\rangle &=& \frac{1}{\sqrt{2}}|0,N,0\rangle\nonumber\\
&&+\frac{1}{\sqrt{2N!}}\left(c_\phi a_1^\dagger -s_\phi a_3^\dagger\right)^N|0,0,0\rangle,
\label{eq-noon}
\end{eqnarray}  
where we define
\begin{eqnarray}
c_\phi &=& \cos\left(\frac{\phi}{2}-\frac{\pi}{4}\right), \quad
s_\phi = \sin\left(\frac{\phi}{2}-\frac{\pi}{4}\right).\nonumber
\end{eqnarray}
Figure \ref{fig:my_label7} presents the change of entanglement entropies $S_1$, $S_2$, and $S_3$ with respect to the parameter $\phi$. The figure clearly shows that the entropy $S_2$ remains constant at $S_2=\log 2$ while the other entropies exhibit a dip at $\phi= \pi/2(3\pi/2)$ with the typical value $S_{1(3)} = \log 2$ of a NOON state.   

\begin{figure}[ht]
    \centering
        \includegraphics[width=8.5cm]{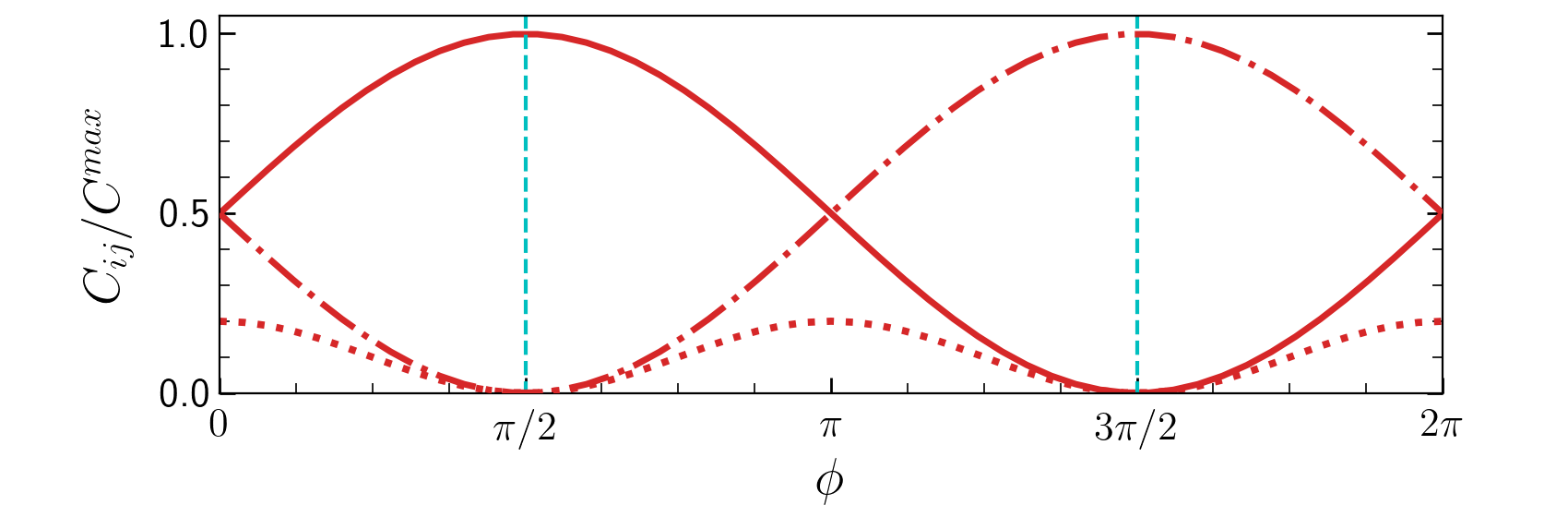}
    \caption{Correlation functions (in units of $C^{max} = N^2/4$) $C_{12}$ (solid line), $C_{23}$ (dot-dashed line) and $C_{13}$ (dot line) for $N=10$ and $U=-1.3$.} 
    \label{fig:my_label8}
\end{figure}

In addition, the two-site correlation functions obtained from the quantum state \textcolor{blue}{(}\ref{eq-noon}\textcolor{blue}{)} are given by (see Figure \ref{fig:my_label8})
\begin{eqnarray}
C_{13}&=&\frac{N(N-2)}{16}\cos^2\phi,\nonumber\\
C_{12}&=&\frac{N^2}{4}\cos^2\left(\frac{\phi}{2}-\frac{\pi}{4}\right),\nonumber\\
C_{23}&=&\frac{N^2}{4}\sin^2\left(\frac{\phi}{2}-\frac{\pi}{4}\right).\nonumber
\end{eqnarray}

From the Figure \ref{fig:my_label8}, it is clear the occurrence of maximum of $C_{12}$ coincides with the cancellation of $C_{23}$ and vice-versa when $C_{13} =0$ at $\phi = \pi/2$ and $\phi = 3\pi/2$, producing the corresponding states
\begin{eqnarray}
|\widetilde{\Psi}_{\text{out}}(\pi/2)\rangle &=&\frac{1}{\sqrt{2}}|0,N,0\rangle +\frac{1}{\sqrt{2}} |N,0,0\rangle,\nonumber\\
|\widetilde{\Psi}_{\text{out}}(3\pi/2)\rangle &=&\frac{1}{\sqrt{2}}|0,N,0\rangle +\frac{(-1)^N }{\sqrt{2}}|0,0,N\rangle,\nonumber
\end{eqnarray}
with the fidelities $F= 0.996123$ and $F= 0.973085$, respectively.
The above NOON states can be seen as the result of an entanglement deconcentration process \cite{Bose1998swapping,Dunningham2002swappingBEC,zhou2013concentrationNOON} through unitary transformation on the NLS state, since they are produced in the subsystems 12 and 23 with less entanglement entropy $S_{1,3}$ than the initial state. This result shows that the protocol controls the transition between the bipartite and tripartite entanglement of the quantum state. It also suggests that the triple-well system can be thought of as a potential shared router operating at the interface between two individual quantum devices to perform a transfer of a NOON state.



\section{Conclusion}

We have proposed a protocol to generate states with controlled levels of entanglement, where the control is realized by breaking the integrability for a short period of time. Our study provides  closed formulas for correlation functions to characterize the entanglement in terms of the integrability breaking time, which allowed us to predict the time required to generate highly entangled states. In the action of protocol on one of the initial states, the maximum correlation predicts the formation of NOON states, whereas, for other unentangled initial states, the maximum correlations are closely related to interference processes.   

Our results have the potential to open new avenues for the manipulation and short-range transfer of entangled states within multimode sytems. These may find applications in quantum routing processes of new devices based on ultracold quantum technology.


\section{Acknowledgments}
The authors acknowledge
support from CNPq (Conselho Nacional de Desenvolvimento Científico e Tecnológico) - Edital Universal 406563/2021-7. 
AF and JL are supported by the Australian Research Council through Discovery Project DP200101339. We thank Rafael Barfknecht for helpful discussions.


\appendix
\section{Interferometry}
In this section, we discuss some interferometric aspects of the protocol proposed in section IV. 

First, we consider the state produced in equation \ref{output1} to calculate the imbalance population between sites 1 and 3. This provides an interference pattern as a function of parameter $\phi$ according to equation
\begin{eqnarray}
\langle  N_1-N_3\rangle\equiv\langle \widetilde{\Psi}_{\text{out}}(\phi)| N_1-N_3|\widetilde{\Psi}_{\text{out}}(\phi)\rangle = -N\cos\phi.\nonumber
\end{eqnarray}
Note the unconventional negative sign can be changed by extending the duration of the last operation to $\Delta t_3 = 3T/4$. The phase uncertainty can be obtained using the error propagation theory~\cite{Dowling2002Rosetta} and is given by 
\begin{eqnarray}
\Delta\phi = \frac{\Delta (N_1-N_3)}{|\partial_\phi\langle N_1-N_3\rangle|} = \frac{1}{\sqrt{N}}.\nonumber
\end{eqnarray}
where the notation $\Delta X =\sqrt{\langle X^2\rangle-\langle X\rangle^2}$ is the standard deviation of operator $X$. The above result shows the uncertainty of parameter $\phi$ is the shot noise limited.

The sensitivity of parameter $\phi$ can be improved for the case of initial twin-Fock state $|l,0,l\rangle$ at sites 1 and 3. This can be achieved by detecting the parity operator $\Pi_1 = e^{-i\pi N_1}$~\cite{Birrittella2021parity}, whose expectation value for the output state generated in equation \ref{output2} is given by
\begin{eqnarray}
\langle \Pi_1\rangle \equiv\langle \widetilde{\Psi}_{\text{out}}(\phi)|\Pi_1|\widetilde{\Psi}_{\text{out}}(\phi) \rangle = P_{l}(\cos(2\phi-\pi)),\nonumber 
\end{eqnarray}
where 
\begin{eqnarray}
P_l(x)=\sum_{k=0}^{\lfloor l/2 \rfloor} \frac{(-1)^k}{2^l} \binom{l}{k}\binom{2l-2k}{l}x^{l-2k},\nonumber
\end{eqnarray}
are the Legendre polynomials. The sensitivity of parameter $\phi$ can be estimated by
\begin{eqnarray}
\Delta \phi = \frac{\Delta \Pi_1}{|\partial_\phi\langle \Pi_1\rangle|},
\end{eqnarray}
which shows the uncertainty of parameter $\phi$ approaches to the Heisenberg limit $\Delta \phi \approx 1/(2l)$ when $\phi \approx \pi/2$ (see \cite{Birrittella2021parity} for details).

\section{Ground state}
\begin{figure}[ht]
    \centering
     \includegraphics[width=8.5cm]{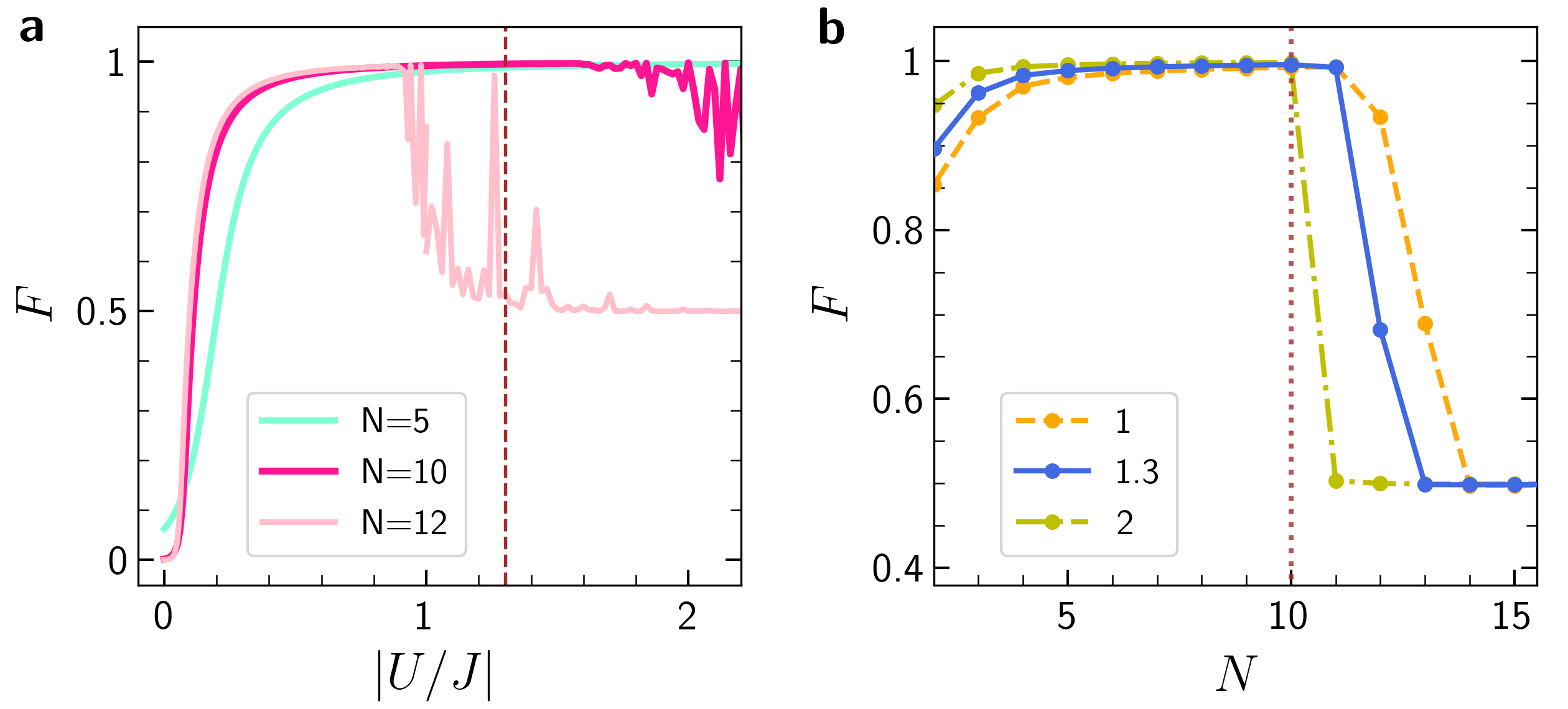}
    \caption{
    (\textbf{a}) Fidelity vs $|U/J|$  for $U<0$ and $N=5$ (turquoise), $N=10$ (magenta) and $N=12$ (pink). (\textbf{b}) Fidelity vs $N$   for $U=-1.0$ (dashed line), $U=-1.3$ (solid line) and $U=-2.0$ (dot dashed line). The vertical lines mark $|U/J|=1.3$ and $N=10$, values used in section \ref{input_cs}.} 
    \label{fig:my_label6}
\end{figure}

In this section, we discuss the structure of the ground state of integrable Hamiltonian (\ref{hint}) in the resonant regime with $U<0$. To this end, we first note that the Hamiltonian  (\ref{hint})  can be reduced to a Bose-Hubbard Hamiltonian of a two-site structure (see, for instance \cite{links2006two})
\begin{eqnarray}
H=U(N_{13}-N_2)^2-J(a_2^\dagger a_{13}+a_{13}^\dagger a_2),\nonumber
\end{eqnarray}
by identifying the single mode operator $\displaystyle a_{13}=\frac{a_1+a_3}{\sqrt{2}}$ and the total number of particles $N_{13}=N_1+N_3$ in the subsystem of sites 1 and 3. On the other hand, for a small number of atoms, it is known that the ground state of two-site Bose-Hubbard Hamiltonian admits the generation of NOON state $\displaystyle |\text{NOON}\rangle = \frac{1}{\sqrt{2}}(|N,0\rangle +|0,N\rangle)$ in strong repulsive interaction regime~\cite{bychek2018noon}, which has the entanglement entropy $S_{\text{NOON}}=-Tr(\rho_1\log\rho_1)=\log 2$ due to it having only one pair of equally likely base Fock state. Likewise, in the resonant regime (with $U < 0$) for small $N\sim 10$, the ground state $|\text{GS}\rangle$ of three modes integrable Hamiltonian (\ref{hint}) presents high fidelity (above 0.99) to the NOON-like state (NLS)
\begin{eqnarray}
|\text{NLS}\rangle = \frac{1}{\sqrt{2}}|0,N,0\rangle+\frac{1}{\sqrt{2}}\frac{(a_{13}^\dagger)^N}{\sqrt{N!}}|0,0,0\rangle.\nonumber
\end{eqnarray}
The Figure \ref{fig:my_label6} presents the fidelity $F = |\langle \text{NLS}|\text{GS}\rangle|^2$ as a function of $|U/J|$ and $N$ for $U<0$. 

\vspace{0.5cm}

\newpage

\bibliographystyle{apsrev4-2}
\bibliography{main}



\end{document}